\documentclass{aa}
\usepackage{psfig,txfonts}

\def\ltsima{$\; \buildrel < \over \sim \;$}
\def\lsim{\lower.5ex\hbox{\ltsima}}
\def\gtsima{$\; \buildrel > \over \sim \;$}
\def\gsim{\lower.5ex\hbox{\gtsima}}

\begin{document}

\title{A look with {\it BeppoSAX} at the low-luminosity \\ Galactic 
X--ray source 4U 2206+54}

\author{N. Masetti\inst{1}
\and
D. Dal Fiume\inst{2,}\thanks{Deceased.}
\and
L. Amati\inst{1}
\and
S. Del Sordo\inst{3}
\and
F. Frontera\inst{1,4}
\and
M. Orlandini\inst{1}
\and
E. Palazzi\inst{1}
}

\institute{
Istituto di Astrofisica Spaziale e Fisica Cosmica -- Sezione di Bologna, 
CNR, via Gobetti 101, I-40129 Bologna, Italy
\and
Istituto Tecnologie e Studio sulla Radiazione Extraterrestre, CNR,
via Gobetti 101, I-40129 Bologna, Italy
\and
Istituto di Astrofisica Spaziale e Fisica Cosmica -- Sezione di
Palermo, CNR, via La Malfa 153, I-90146 Palermo, Italy
\and
Dipartimento di Fisica, Universit\`a di Ferrara, via Paradiso 12, I-44100
Ferrara, Italy
}

\titlerunning{A look with {\it BeppoSAX} at 4U 2206+54}
\authorrunning{Masetti et al.}

\offprints{N. Masetti, {\tt masetti@bo.iasf.cnr.it}}

\date{Received February 16, 2004; accepted April 16, 2004}

\abstract{A pointed observation of the low-luminosity galactic source 
4U 2206+54 was carried out in November 1998 with {\it BeppoSAX}. The light 
curve of 4U 2206+54 shows erratic variability on a timescale of $\sim$1 
hour; neither hardness variations nor time periodicities are detected 
throughout this 67 ks long observation. Thanks to the wide spectral 
coverage capabilities of {\it BeppoSAX} we could observe the source X--ray 
continuum over three energy decades, from 0.6 to 60 keV. 
The spectrum could be equally well fitted either with a blackbody plus 
Comptonization or with a high energy cutoff power law. No iron emission 
around 6.5 keV was detected, while a tentative detection of a cyclotron 
resonant feature in absorption is presented. Comparison of the present 
{\it BeppoSAX} data with the information available in the literature for 
this source suggests that 4U 2206+54 is a close binary system in which a 
(possibly magnetized) NS is accreting from the companion star wind.

\keywords{Stars: binaries: close --- X--rays: binaries --- Stars:
neutron --- Stars: individuals: 4U 2206+54 --- Accretion, accretion 
disks}}

\maketitle

\section{Introduction}

Massive X--ray Binaries (MXRBs) are double systems composed of a compact
object, generally a neutron star (NS), orbiting an early-type star and
accreting matter from it. 
In X--rays, MXRBs can be seen as persistently bright, with luminosities 
greater than 10$^{35}$ erg s$^{-1}$, or present a transient behaviour
characterized by quiescent phases, with emissions around 10$^{34}$ erg
s$^{-1}$ or less, followed by intense (up to
$\sim$10$^{38}$ erg s$^{-1}$ at peak) outbursts; in several cases,
these outbursts show a periodic trend as a result of the orbital motion of
the NS along a highly eccentric orbit (see White et al. 1995 and van
Paradijs 1995 for a review). 
Usually, the former group is associated with compact objects
steadily accreting from the companion via Roche lobe overflow and/or
stellar wind, while in the latter one accretion is discontinuous and
occurs when the compact source enters a disk-like envelope around the
companion star or, more generally, interacts more closely with the 
companion as it approaches periastron (e.g., Corbet 1986).

There are however MXRBs which do not fit this classification, i.e. the
so-called low-luminosity MXRBs, characterized by their relatively low
persistent emission in the X--ray domain (10$^{34}$--10$^{35}$ erg
s$^{-1}$) compared to those of persistent MXRBs and which do not display
outbursts. The X--ray source 4U 2206+54 is one of these objects.

It was discovered by {\it Uhuru} (Giacconi et al. 1972), was
monitored with {\it EXOSAT} between 1983 and 1985 by Saraswat \& Apparao
(1992) who reported aperiodic hard flares from the source lasting a few
hundred seconds and producing variations of a factor 3 to 5, and long-term
variations of a factor of $\sim$20 in the 2--10 keV persistent luminosity
($\approx$0.3--5$\times$10$^{34}$ erg s$^{-1}$). These authors 
also reported a pulse period of $\sim$400 s, which however has been 
recently questioned by Corbet \& Peele (2001) from a re-analysis of the 
{\it EXOSAT} data as well as of archival {\it RXTE} data. Corbet \& Peele 
(2001) further reported, on the basis of ASM observations, a 9.6-d 
periodicity in the X--ray flux of the source; they also modeled the 
{\it RXTE} spectra obtained on two occasions (March 11 and 13, 1997) using 
a power law modified with an exponential cutoff. They found a flux 
decrease by a factor of three (from 3.12$\times$10$^{-10}$ to 
1.14$\times$10$^{-10}$ erg cm$^{-2}$ s$^{-1}$) between the two pointings. 
Negueruela \& Reig (2001) reanalyzed the {\it RXTE} pointing of March 11, 
1997 obtaining comparable results; they also confirmed the presence of
flares during which they found a positive correlation between source
hardness and flux. These authors also did not detect any X--ray pulsation
from the object.

The X--ray spectral characteristics of this source are typical of
accretion onto a NS from a wind coming from the companion star (Negueruela
\& Reig 2001), although the hypotheses of an accreting white dwarf
(WD: Saraswat \& Apparao 1992; Corbet \& Peele 2001) or black hole 
(BH: Negueruela \& Reig 2001) were also considered. 

The optical counterpart was identified by Steiner at al. (1984) as the
early-type star BD +53$^{\circ}$2790, located at 2.5 kpc from the Earth.
This star was subsequently thoroughly studied in its optical-UV
spectroscopic properties by Negueruela \& Reig (2001) who classified it as
a peculiar late O-type active star. No radio counterpart has been
detected so far (Nelson \& Spencer 1988).

The information available on 4U 2206+54 is therefore not conclusive
to understand the nature of the accreting source. In particular, the
lack of knowledge of its X--ray spectrum above 30 keV hinders any
hypothesis on the presence of a Cyclotron Resonant Feature (CRF) and
thus any conjecture on the magnetic field of the accreting source as well
as on its nature. Likewise, the poor sampling concerning the soft side of 
the X--ray spectrum never allowed an accurate estimate of the hydrogen 
column density $N_{\rm H}$; also, this did not allow a sensitive search 
for a soft component in the emission from this source.
Moreover, a further independent check of the presence of a periodic
variability (or the lack thereof) in X--rays is also needed.

Therefore, to explore the timing and spectral behaviour of 4U 2206+54 over
a broad spectral range, with particular attention to both soft ($<$2 keV)
and hard ($>$20 keV) X--ray domains, we observed this source with {\it
BeppoSAX} (Boella et al. 1997a).

The paper is organized as follows: Sect. 2 will illustrate the
observations and the data analysis, while in Sect. 3 the results
showing the X--ray spectral and timing behaviours of 4U 2206+54 will be
reported; in Sect. 4 a discussion will be given.

\section{The {\it BeppoSAX} pointing}

4U 2206+54 was observed between November 23 and 24, 1998, for a total
on-source time of $\sim$67 ks. 
The source was observed with three of the four coaligned Narrow-Field
Instruments (NFIs) mounted on {\it BeppoSAX}: the Low Energy Concentrator
Spectrometer (LECS, 0.1--10 keV; Parmar et al. 1997), two Medium Energy
Concentrator Spectrometers (MECS, 1.5--10 keV; Boella et al. 1997) and 
the Phoswich Detection System (PDS, 15--300 keV; Frontera et al. 1997).
The High-Pressure Gas Scintillation Proportional Counter (HPGSPC, 6--60
keV; Manzo et al. 1997) was temporarily not available between November 18
and 25, 1998: therefore, no data for 4U 2206+54 were obtained with this 
instrument. The total duration of this {\it BeppoSAX} pointing along with 
the on-source exposure times for each used NFI are reported in Table~1.

\begin{table}
\caption[]{Log of the {\it BeppoSAX} observation presented in this
paper.}
\begin{center}
\begin{tabular}{lccccc}
\noalign{\smallskip}
\hline
\noalign{\smallskip}
\multicolumn{1}{c}{Start day} & Start time & Duration & 
\multicolumn{3}{c}{On-source time (ks)} \\
 & (UT) & (ks) & LECS & MECS & PDS \\
\noalign{\smallskip}
\hline
\noalign{\smallskip}
 1998 Nov 23 & 16:02:49 & 66.6 & 12.6 & 33.6 & 15.2 \\
\noalign{\smallskip}
\hline
\noalign{\smallskip}
\end{tabular}
\end{center}
\end{table}

Good NFI data were selected from intervals outside the South Atlantic
Geomagnetic Anomaly when the elevation angle above the earth limb was
$>$$5^{\circ}$, when the instrument functioning was nominal and, for LECS
events, during spacecraft night time. The SAXDAS 2.0.0 data analysis
package (Lammers 1997) was used for the extraction and the processing of
LECS and MECS data. The PDS data reduction was instead performed
using XAS version 2.1 (Chiappetti \& Dal Fiume 1997). 
LECS and MECS data were reduced using an extraction radius of 6$'$
and 4$'$, respectively, centered at the source position; before
extraction, data from the two MECS units were merged.
Background subtraction for the two imaging instruments was performed using
standard library files, while the background for the PDS data was
evaluated from the fields observed during off-source pointing intervals.

Because 4U2206+54 is located near the Galactic plane and is not
a particularly bright X-ray binary, we checked for possible effects
induced by Galactic diffuse emission in the PDS data background
evaluation. The off-source fields for background evaluation were indeed
at different Galactic latitudes ($b$ = $-$4$\fdg$0 and $b$ = +1$\fdg$6) 
with respect to the source ($b$ = $-$1$\fdg$2), so a gradient in the 
Galactic diffuse emission could potentially be present. However, the count 
rate difference between the two fields is 0.07$\pm$0.06 counts s$^{-1}$, 
thus consistent with zero; moreover, this difference impacts on the 
background estimate by less than 0.5\%. Therefore we considered this
effect negligible.

\begin{figure*}[t!]
\psfig{figure=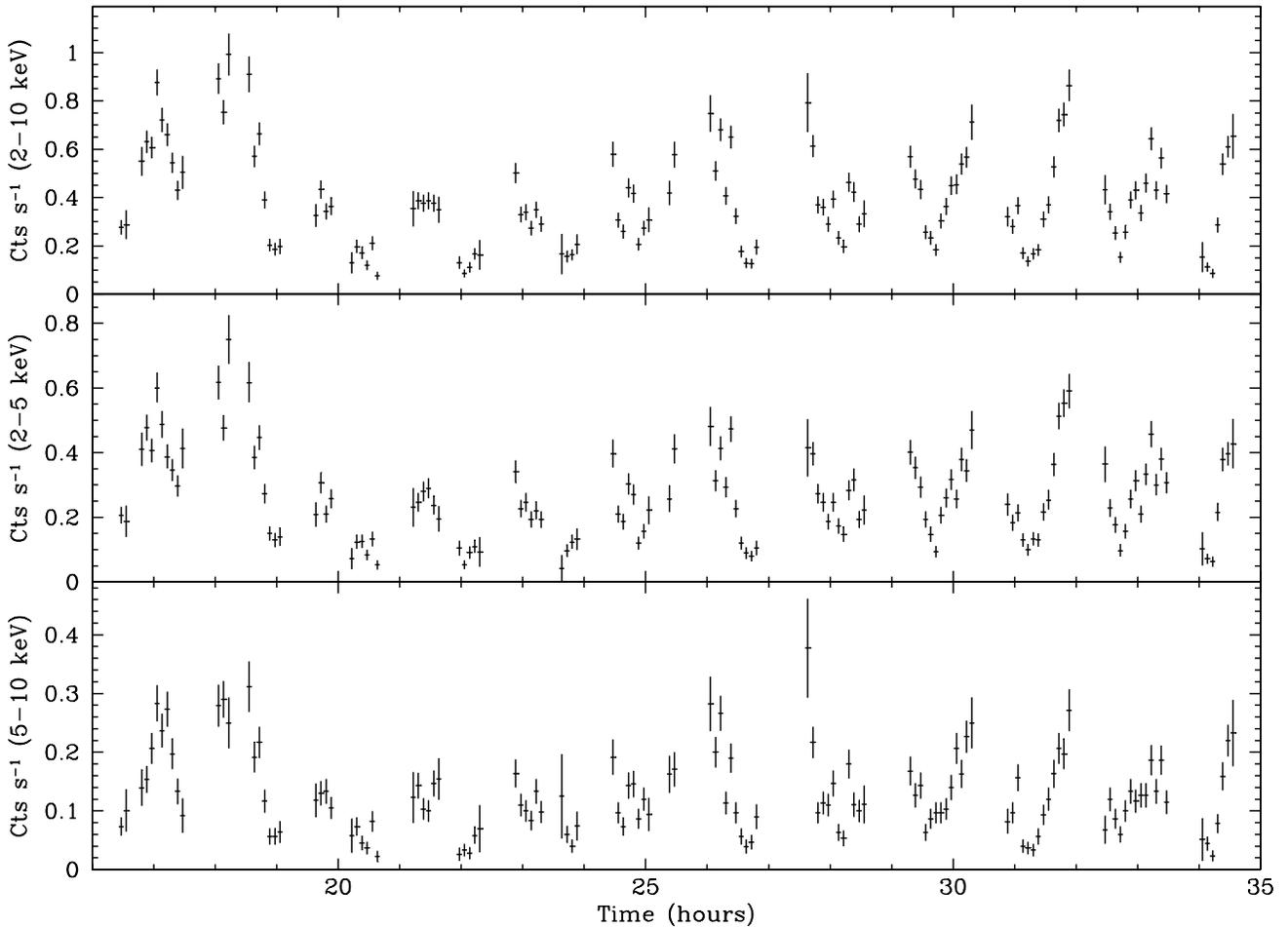,width=18cm,angle=-90}
\caption[]{Background-subtracted X--ray light curves of 4U 2206+54 in the 
2--10 keV {\it (upper panel)}, 2--5 keV {\it (middle panel)} and 5--10 keV 
{\it (lower panel)} ranges as observed during the {\it BeppoSAX} pointing.
All curves are rebinned at 300 s. Flaring activity on timescales 
of $\sim$1 hour can be noticed. Times are expressed in hours starting 
from 00:00 UT of November 23, 1998. Vertical error bars show 1$\sigma$ 
confidence level uncertainties for each bin.}
\end{figure*}

\section{Results}

\subsection{Light curves and timing analysis}

The 2--10, 2--5 and 5--10 keV MECS light curves of 4U 2206+54
as seen during the {\it BeppoSAX} pointing, background-subtracted and 
rebinned at 300 s, are displayed in Fig. 1.
They show substantial random variability (up to a factor $\sim$10 overall)
in the form of repeated flares lasting $\sim$1 hr with internal
fluctuations down to timescales of $\sim$50--100 s.

To see if this erratic behaviour implied spectral changes depending on the
source intensity, we computed a hardness ratio between the 5--10 keV and
the 2--5 keV count rates and plotted it against the total 2--10 keV count
rate. The results are shown in Fig. 2: apparently, there is no dependence
of the hardness ratio on the total intensity of 4U 2206+54 in the 2--10
keV range. This result differs from the findings of Negueruela \& Reig
(2001) who found that the source became harder with increasing X--ray
intensity during an {\it RXTE} pointing.

Timing analysis on the 2--10 keV data (i.e. where the S/N was highest)
was performed with the FTOOLS v5.1\footnote{available at: \\ {\tt
http://heasarc.gsfc.nasa.gov/ftools/}} (Blackburn 1995)
tasks {\tt powspec} and {\tt efold}, after having converted the event
arrival times to the solar system barycentric frame.
The results do not reveal the presence of any kind of periodicity or
quasiperiodic oscillation. In particular, we did not detect
the 392 s periodicity reported by Saraswat \& Apparao (1992), thus
confirming the negative findings of Corbet \& Peele (2001) and Negueruela
\& Reig (2001). We get a 90\% confidence level upper limit of $\sim$8\%
in the amplitude of the signal in the 2--10 keV light curve induced by the
above periodicity, consistent with the results of Corbet \& Peele (2001).

The Power Spectral Density (PSD) obtained with this analysis is
characterized by red noise and shows no significant deviations from the
shot-noise behaviour, similarly to that found by Negueruela \& Reig
(2001). The $rms$ variability of the 2--10 keV light curve is $\sim$40\%
in the 10$^{-3}$--1 Hz range, with no significant differences when only 
the softer (2--5 keV) or the harder (5--10 keV) band is considered, in
agreement with the result, presented above, that no spectral dependence as
a function of the total source intensity is found.

\begin{figure}[t!]
\psfig{figure=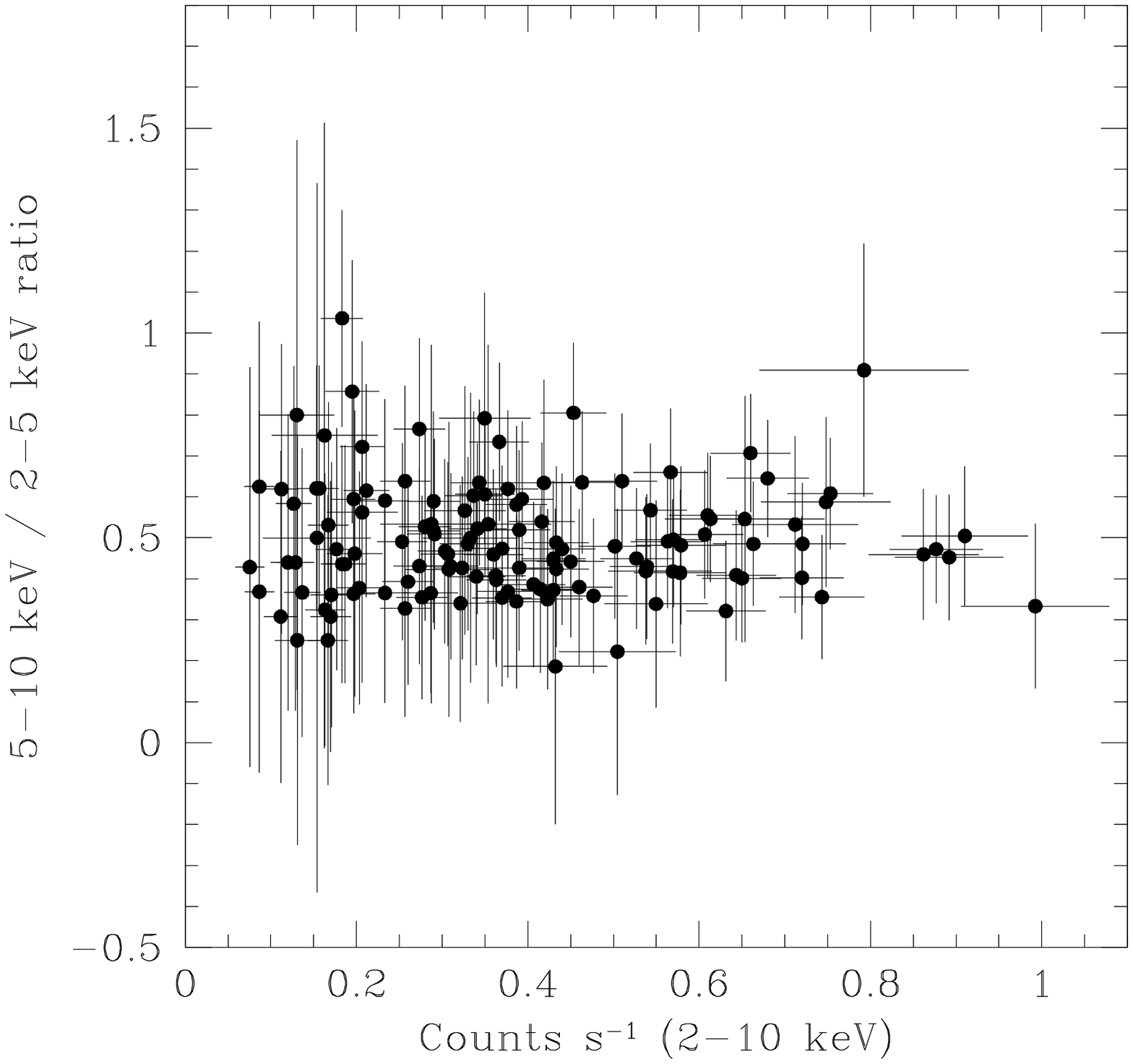,width=8.8cm,angle=0}
\caption[]{Hardness-intensity diagram for 4U 2206+54 during the {\it 
BeppoSAX} pointing. No significant trend of the 5--10 keV / 2--5 keV
hardness ratio with increasing 2--10 keV intensity is found. Error bars 
show 1$\sigma$ confidence level uncertainties for each point.}
\end{figure}

\subsection{Spectra}

\begin{table*}
\caption[]{Best--fit spectral parameters for 4U 2206+54. Quoted
errors are at 90\% confidence level for a single parameter. Quantities 
in square brackets are frozen at the indicated value. Luminosities,
corrected for interstellar Galactic absorption, are computed assuming a
distance $d$ = 2.5 kpc and are expressed in units of 10$^{34}$ erg
s$^{-1}$.}
\begin{center}
\begin{tabular}{c|cccc}
\hline
\noalign{\smallskip}
 & (1) & (2) & (3) & (4) \\
\noalign{\smallskip}
\hline
\noalign{\smallskip}
Parameter & PL+{\sc HighEcut} & {\sc BrokenPL} & BB+{\sc compST} & 
BB+{\sc compST}+CRF\\
\noalign{\smallskip}
\hline
\noalign{\smallskip}
$\chi^{2}$/dof & 245.0/219 & 233.3/219 & 228.9/218 & 225.0/217 \\

$N_{\rm H}$$^{a}$ & 0.88$^{+0.21}_{-0.19}$ & 0.86$^{+0.20}_{-0.16}$ &
0.81$\pm$0.21 & 0.77$^{+0.20}_{-0.18}$ \\ 

$\Gamma$ & 0.95$^{+0.11}_{-0.14}$ & 0.99$^{+0.10}_{-0.12}$ & --- & --- \\

$\Gamma_2$ & --- & 2.03$^{+0.12}_{-0.11}$ & --- & --- \\

$E_{\rm cut}$ (keV) & 4.3$^{+0.6}_{-0.5}$ & 5.6$\pm$0.4 & --- & --- \\

$E_{\rm fold}$ (keV) & 10.6$^{+2.7}_{-2.0}$ & --- & --- & --- \\

$N_{\rm PL}$ ($\times$10$^{-3})$ & 3.6$\pm$0.6 & 3.7$^{+0.7}_{-0.5}$ &
--- & --- \\

$T_{\rm BB}$ (keV) & --- & --- & 1.63$^{+0.12}_{-0.16}$ &
1.61$^{+0.10}_{-0.17}$ \\

$R_{\rm BB}$ (km) & --- & --- & 0.15$^{+0.03}_{-0.02}$ & 0.15$\pm$0.02 \\

$T_{\rm e^-}$ (keV) & --- & --- & 9$^{+9}_{-3}$ & 19$^{b}$ \\

$\tau$ & --- & --- & 9$^{+5}_{-3}$ & 7$^{+3}_{-4}$ \\

$N_{\rm Comp}$ ($\times$10$^{-3}$) & --- & --- & 3.3$\pm$1.1 &
3.0$^{+1.0}_{-0.8}$ \\

\noalign{\smallskip}
\hline
\noalign{\smallskip}

$E_{\rm CRF}$ (keV) & --- & --- & --- & [35] \\

$W_{\rm CRF}$ (keV) & --- & --- & --- & [10] \\

$\delta_{\rm CRF}$ & --- & --- & --- & 1.4$^{+1.6}_{-1.1}$ \\

\noalign{\smallskip}
\hline
\noalign{\smallskip}
\hline
\noalign{\smallskip}
$L_{\rm 0.5-2~keV}$$^{c}$ & 0.65 & 0.67 & 0.67 & 0.63 \\
$L_{\rm 2-10~keV}$        & 3.15 & 3.10 & 3.09 & 3.09 \\
$L_{\rm 10-50~keV}$       & 3.01 & 3.95 & 3.74 & 3.50 \\

\noalign{\smallskip}
\hline
\hline
\multicolumn{5}{l}{$^{a}$ In units of 10$^{22}$ cm$^{-2}$} \\
\multicolumn{5}{l}{$^{b}$ Poorly constrained} \\
\multicolumn{5}{l}{$^{c}$ This value was computed by extrapolating the
best-fit models down to 0.5 keV} \\
\end{tabular}
\end{center}
\end{table*}

\begin{figure}[t!]
\psfig{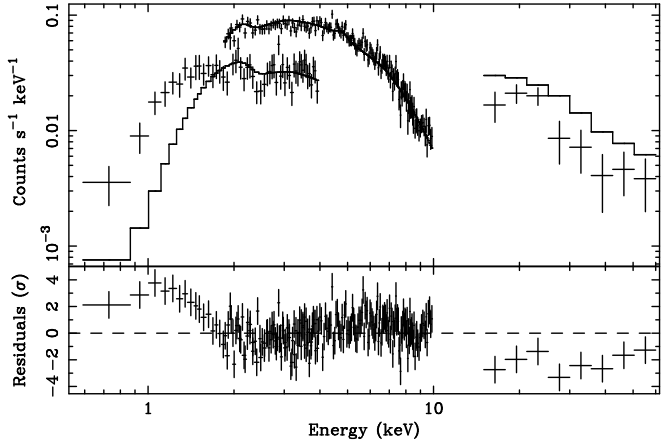}
\quad
\caption[]{Average 0.6--60 keV X--ray spectrum of 4U 2206+54 obtained with
the {\it BeppoSAX} NFIs and fitted with a photoelectrically absorbed PL.
The best-fit model is shown as a continuous line. Clearly, this model
(with a reduced $\chi^2$ = 1.7) does not provide an acceptable
description of the spectral data, particularly below 2 keV and above 10
keV, as seen in the panel reporting the residuals.}
\end{figure}

To perform spectral analysis, the NFI pulse-height spectra were rebinned 
to oversample by a factor of 3 the full width at half maximum
(FWHM) of the energy resolution and to have a minimum of 20 counts per
bin, such that $\chi^2$ statistics could reliably be used.
Data were then selected, for each NFI, in the energy intervals in which 
sufficient counts were detected from the source and for which the 
instrument response function was well determined. This led us to consider
the spectral interval 0.6--4 keV for the LECS, 1.8--10 keV for the MECS
and 15--60 keV for the PDS. We then used the package {\sc xspec} (Dorman
\& Arnaud 2001) v11.0 to fit the resulting broad band energy spectrum.

We included in all fits described here an interstellar photoelectric
absorption column, modeled using the Wisconsin cross sections as 
implemented in {\sc xspec} (Morrison \& McCammon 1983) and with solar
abundances as given by Anders \& Grevesse (1989).

When performing the spectral fits, normalization factors were applied
to LECS and PDS spectra following the cross-calibration tests between
these instruments and the MECS (Fiore et al. 1999). These factors were
constrained to be within the allowed ranges during the spectral fitting.

The {\it BeppoSAX} data (Figs. 3 and 4) clearly show continuum emission 
both above 30 keV and below 1 keV. To the best of our knowledge this is 
the first time that X--ray emission outside the 1--30 keV range is 
reported from this source. 

Motivated by the results obtained in Sect. 3.1, and to further
test the dependence (or lack thereof) of the 4U 2206+54 spectral shape on
the total source intensity, we created two time windows: one in which the
source was above 0.3 counts s$^{-1}$ in the 2--10 keV band (`high state')
and the other characterized by the source being below 0.3 counts s$^{-1}$
(`low state') in this energy range.  We then accumulated the spectra over
these two windows and compared them by using the best-fit models in Table
2 (see the analysis of time-averaged spectra below). As expected, given
the results in Sect. 3.1, no significant parameter variations (with of
course the exception of the model normalizations) were found between these
high- and low-state spectra of 4U 2206+54 for any of the tested best-fit
models.

We therefore considered the average spectrum of 4U 2206+54 given that no
hints of variation in the spectral shape during the {\it BeppoSAX}
pointing were suggested by the inspection of the color-intensity diagram
of the source (reported in Fig. 2). Table 2 reports the models and
parameters thereof that best fit the spectral data. Figure 5 shows
instead the data residuals associated with each of the best-fit model
presented in Table 2. 

A possible concern in considering the average spectrum might rise from the
fact that, during the {\it BeppoSAX} observation, the source was strongly
variable; the NFIs may thus have sampled different subsets of the light
curve during the pointing. This could have introduced spurious spectral
distortions. However, given that the intercalibration factors among the
three instruments were in the allowed ranges even when they were left free
to vary during the spectral fits, we are confident that this did not 
produce any effect on the averaged {\it BeppoSAX} spectrum.

We first tried to fit the overall 0.6--60 keV spectrum of 4U 2206+54
detected by the {\it BeppoSAX} NFIs by using a simple power law (PL).
The results, as shown in Fig. 3, are clearly unsatisfactory, with a
reduced $\chi^2$ = 1.7, an excess at low energies below 2 keV and a 
deficit at high energies above 10 keV. So we applied the phenomenological 
model made of a PL modified by a high-energy exponential cutoff. This
is considered the ``classical" way of describing the X--ray spectra
of MXRBs (e.g., White et al. 1983a). As in Negueruela \& Reig (2001), an
acceptable fit -- see column (1) of Table 2 -- was obtained.

A similarly good fit, as reported in column (2) of Table 2, was also 
achieved using a broken PL. Other simple models such as blackbody (BB), 
disk-blackbody (DBB; Mitsuda et al. 1984) and thermal bremsstrahlung (TB) 
provided instead poor fits. 

We then tried to fit the spectrum by using a more physical description. 
Given that simple models failed, we tried a composition of two models. 
The best results were obtained considering a BB plus a Comptonization 
model. For the latter, we used the first-order modelization by Sunyaev
\& Titarchuk (1980; {\sc compST} in {\sc xspec}). In this case also, an 
acceptable fit was achieved; this case is shown in Fig. 4, and its 
parameters are in column (4) of Table 2. The fit a DBB instead of a BB
makes the {\sc compST} parameter $\tau$ unconstrained.

A description of the {\it BeppoSAX} spectrum made by using the more 
accurate and complex Comptonization model by Titarchuk (1994; {\sc compTT} 
in {\sc xspec}), with either the addition of a BB or not, failed to
provide a sound value of $N_{\rm H}$: this was always unacceptably (i.e.,
up to two orders of magnitude) lower than the Galactic one, which is
$N_{\rm H}^{\rm G}$ = 0.59$\times$10$^{22}$ cm$^{-2}$ according to Dickey
\& Lockman (1990); this model could not constrain all the Comptonization 
(and BB, if present) parameters.
Nonetheless, the best-fit values of the {\sc compTT} parameters were 
consistent with those obtained from the Comptonization component 
in the best-fit BB+{\sc compST} model listed in Table 2: this suggests 
that the failure of the {\sc compTT} model is more likely due to its
complexity compared to the overall quality of the spectrum, rather than 
to a completely wrong description of the source spectral shape.
In this case also, the addition of a DBB to the {\sc compTT} makes the 
model even more unstable.

We also performed a fit of the data with the {\sc xspec} models {\sc
raymond}, {\sc mekal} and {\sc vmekal}, which describe the emission from a
hot diffuse gas (Raymond \& Smith 1977; Mewe et al. 1985). All these
descriptions provided a poor description of the overall spectral shape
(that is, returning a reduced $\chi^{2} \sim$ 1.5 in all cases), as
already noted by Negueruela \& Reig (2001) from the analysis of {\it RXTE}
data. 
We note that, by adding a BB with $kT_{\rm BB} \sim$ 1.6 keV to any of
these models, the fit becomes formally acceptable; in this case, however,
the temperature of the emitting diffuse gas is uncomfortably large ($>$60
keV) and unconstrained. So we consider this spectral modelization not
viable. A similar result was obtained when considering a BB+TB model. 

An attempt to use complex absorbers to describe the low-energy part
of the spectrum was also made, given that 4U 2206+54 may be a system in
which part of the absorption is intrinsic (Negueruela \& Reig 2001).
However, fits performed with a partial covering absorber did not
improve the spectral description and did not allow us to constrain either
the additional $N_{\rm H}$ or the covering percentage (around
5-6$\times$10$^{22}$ cm$^{-2}$ and 60-65\%, respectively).
The use of an ionized absorber instead of a neutral one did not produce
any significant difference in the fit results. These findings were
practically independent of the chosen basic model (Comptonization or PL).

\begin{figure*}[t!]
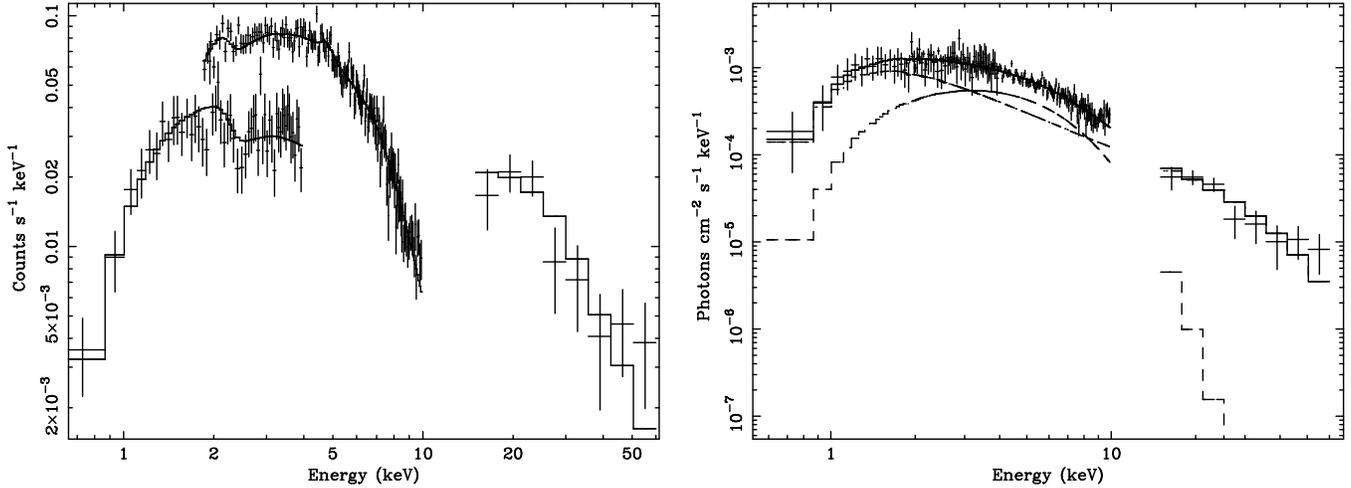

\centering{\mbox{\psfig{figure=0273f4l.ps,width=8.7cm,angle=-90}}}
\quad
\psfig{figure=0273f4r.ps,width=8.7cm,angle=-90}
\caption[]{{\it Left panel}: average 0.6--60 keV X--ray spectrum of 4U 
2206+54 obtained with the {\it BeppoSAX} NFIs and fitted with a 
photoelectrically absorbed BB plus Comptonization. The best-fit model is 
shown as a continuous line. {\it (Right panel)}: unfolded {\it BeppoSAX} 
photon spectrum fitted with this same model. The two components are 
indicated with short-dashed (BB) and long-dashed (Comptonization) lines, 
while the overall model is again indicated with a continuous line.}
\end{figure*}

No evidence for an iron emission line around 6.5 keV was found: we
obtained upper limits (at 90\% confidence level) to the equivalent width
of $<$56 eV, $<$65 eV and $<$156 eV to the equivalent width in case of a
line with narrow (0.1 keV), broad (0.5 keV) and very broad (1 keV) FWHM, 
respectively. These values are largely independent of the chosen best-fit 
model.

Following the suggestion by Corbet \& Peele (2001) we then tried to look 
for a signature of a highly magnetic field NS, through the search for a
CRF in absorption, in the high-energy part of the {\it BeppoSAX}
spectrum. To do this, we used the {\sc cyclabs} multiplicative model
(Mihara et al. 1990; Makishima et al. 1990) in {\sc xspec}.
The limited S/N ratio above 15 keV prevented us from fully exploring 
this possibility; however, to limit the number of free parameters
introduced by this addition to the spectrum description, we chose 
to freeze the line energy and width. The latter was fixed to a value
typical of magnetic NSs in MXRBs, i.e. 10 keV (e.g. Orlandini \& Dal
Fiume 2001), while we made the former vary between 10 and 60 keV in steps
of 5 keV among the fits.
We found that only in the case of the BB+{\sc compST} model a CRF with a 
best-fit line energy of 35 keV was justified -- see column (4) of 
Table 2 --, albeit at a 2$\sigma$ confidence level only, according to the 
results obtained by running an F-test. The other best-fit models in Table 2
(broken PL and high-energy cutoff PL) are not improved by the addition of
a CRF. Thus, given all the assumptions and the uncertainties described
above, we regard this CRF identification as tentative only. 

Assuming a distance to 4U 2206+54 of 2.5 kpc (Steiner et al. 1984) and
using the models described above for the average spectrum of the
source, we can evaluate the unabsorbed luminosities of the source in the
0.5--2 keV, 2--10 keV and 10--50 keV bands for each best-fit model. These
values are reported in Table 2. The ratios between the BB and
Comptonization contributions in the BB+{\sc compST} model in the three 
X--ray bands above are 0.17, 1.98 and 0.07, respectively.

\begin{figure}
\psfig{figure=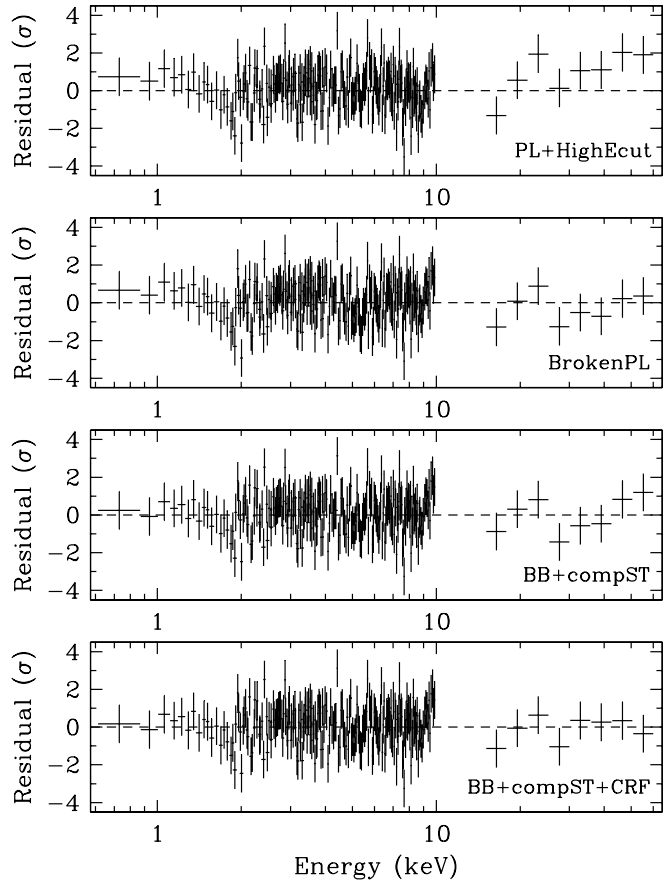,width=9.2cm,angle=0}
\caption[]{Comparison among the residuals obtained with the best-fit 
models listed in Table 2. Names associated with each panel refer to 
columns (1)--(4) of this Table.}
\end{figure}

\section{Discussion}

The {\it BeppoSAX} observation of 4U 2206+54 reported in this paper
allowed us to study the properties of the source in great detail, and for
the first time, in the 0.6--60 keV range. The source exhibited an average
2--10 keV unabsorbed luminosity of 3.1$\times$10$^{34}$ erg s$^{-1}$ during 
this pointing, thus comparable with the ``high-state" emission observed by
Saraswat \& Apparao (1992) in 1983 and 1985 from {\it EXOSAT} data, but
$\sim$10 times brighter than that observed by these authors in 1984, and a
factor $\sim$10 fainter than during the 1997 {\it RXTE} pointing (e.g.,
Negueruela \& Reig 2001). Thus the source can vary by about 10 fold, and
as much as a factor of $\sim$100, on year-long timescales. This can also be
seen in Fig. 1 of Corbet \& Peele (2001), where the entire {\it RXTE}/ASM
1.5--12 keV light curve of 4U 2206+54 between years 1996 and 2001 is
shown. The periodogram in Fig. 2 of these authors shows a peak at a 
frequency of $\approx$4$\times$10$^{-3}$ d$^{-1}$: if real, this would 
indicate the presence of a further (superorbital?) periodicity of
$\approx$250 d in the system.

On much shorter time scales, the 2--10 keV light curve shows flares of
remarkable (a factor ten) intensity variations lasting about 1 hr, with
finer variability down to timescales of 50--100 s. This flaring emission
does not appear to imply spectral variation correlated with the intensity.
Moreover, the PSD analysis indicates that the variable X--ray emission
from the source is due to a stochastic phenomenon. These long- and
short-term variability characteristics point to an explanation for this
X--ray activity as due to random inhomogeneities in the accretion flow
onto a compact object (e.g. van der Klis 1995).

The colorless variability result obtained from our {\it BeppoSAX} data is
at odds with that found by Saraswat \& Apparao (1992) and by Negueruela \&
Reig (2001). The explanation for this is not clear. Concerning the
findings by Saraswat \& Apparao (1992) this discrepancy can be due to
slight secular variations of the source hardness ratio in observations
separated in time by about two years: indeed, if one considers the 1983
and 1985 {\it EXOSAT} data in their Fig. 4 separately, no significant
hardness-intensity dependence is observed. As regards instead the 
discrepancy with the data from the {\it RXTE} pointing in Negueruela \& 
Reig (2001) it may be possible that this hardness-intensity dependence 
becomes more evident at larger source luminosities.

Moreover, the PSD of {\it BeppoSAX} data does not show any periodicity in
the 10$^{-3}$--1 Hz range. We thus independently confirm the results by
Corbet \& Peele (2001) and Negueruela \& Reig (2001) that no $\approx$400
s X--ray modulation comes from 4U 2206+54.

X--ray spectroscopy with {\it BeppoSAX} can help us in better 
understanding the accretion dynamics as well as the nature of the 
accretor. Indeed, we obtained for 4U 2206+54 an unprecedented simultaneous 
spectral coverage of the 0.6--60 keV range. Our results show that 
the ``classic" model generally used to fit the X--ray spectra of MXRBs 
hosting an accreting NS (White et al. 1983a) works very well, as in the 
case of the March 1997 {\it RXTE} data. In these observations the spectral 
parameters $N_{\rm H}$, $\Gamma$, $E_{\rm cut}$ and $E_{\rm fold}$ were in
the range 1.12--1.71, (2.7--4.6)$\times$10$^{22}$ cm$^{-2}$, 5.3--7.3 and
10.5--17.3, respectively, with the values at the lower edge of the
interval holding at low source fluxes (Corbet \& Peele 2001).

However, when we compare our best-fit parameters with the {\it RXTE}
spectral findings obtained when the source flux was highest (corresponding
to a 2--10 keV luminosity of 2.3$\times$10$^{35}$ erg s$^{-1}$), we see
that in our data (i) the spectral slope $\Gamma$ is substantially harder;
(ii)  both characteristic energies of the model, $E_{\rm cut}$ and $E_{\rm
fold}$, are lower by a factor $\sim$2; (iii) the hydrogen column density
$N_{\rm H}$ is lower by a factor $\sim$6. Comparison with the
lower-intensity {\it RXTE} observation (which implies a 2--10 keV
luminosity of 8.5$\times$10$^{34}$ erg s$^{-1}$) shows a consistency with 
the {\it BeppoSAX} data barring the $N_{\rm H}$ value, which is still 
$\sim$3 times higher in the {\it RXTE} results.

The results of points (i) and (ii) above may reflect
an actually different spectral shape due to a different emission level
from the source with respect to that observed in the March 1997 {\it RXTE}
data. Indeed, Table 1 of Corbet \& Peele (2001) shows that $\Gamma$ (which 
they indicate as $\alpha$ in their paper), $E_{\rm cut}$ and $E_{\rm 
fold}$ appear to inversely correlate with the source flux, and a simple 
computation indicates that the source gets harder as the flux increases.
Therefore, it appears that there is a switch to harder source spectra when 
4U 2206+54 overcomes a threshold luminosity with a value lying somewhere 
between 1$\times$10$^{35}$ and 2$\times$10$^{35}$ erg s$^{-1}$. 
Unfortunately, in the light of the above data, we cannot say if this 
transition occurs smoothly or in the form of a ``parameter jump".

Instead, we believe that the $N_{\rm H}$ measurement obtained with {\it
BeppoSAX} is substantially different and more reliable thanks to the better
spectral coverage at low energies afforded by the {\it BeppoSAX} LECS.
Indeed, if we consider only our data above 2.5 keV, we obtain that $N_{\rm
H}$ = 2.5$^{+0.9}_{-1.0}$ $\times$10$^{22}$ cm$^{-2}$. Thus, in our
opinion the $N_{\rm H}$ value obtained with {\it BeppoSAX} should be
considered as the correct one. We note that the order-of-magnitude 
difference in flux between the {\it BeppoSAX} and the {\it RXTE} 2--10 keV 
measurements cannot be explained by the different $N_{\rm H}$ estimate, 
which at most may account for a flux difference of about 30\% only.

This new value of $N_{\rm H}$, (0.8--0.9)$\times$10$^{22}$ cm$^{-2}$),
compares much better with the Galactic value along the 4U 2206+54 line of
sight (0.59$\times$10$^{22}$ cm$^{-2}$) and with the optical $V$-band
reddening value ($A_V$ = 1.6) given by Negueruela \& Reig (2001):  
although the empirical relation between $A_V$ and $N_{\rm H}$ by Predehl
\& Schmitt (1995) implies the presence of further hydrogen local to the
source, the difference is now to within a factor of two, and not an order 
of magnitude as from the previous $N_{\rm H}$ estimates. This alleviates 
the conundrum, stressed by Negueruela \& Reig (2001), of the non-detection 
of iron emission in presence of very optically thick material around the
X--ray source. Concerning this issue, our observation allowed us to put
tighter upper limits on the presence of any X--ray Fe emission with
respect to the one determined by {\it EXOSAT} (Gottwald et al. 1995).

The spectrum description by means of a more physically sound model, 
namely a BB+Comptonization, again points to the presence of a very 
compact object as the accretor: the spectral shape and the temperatures of 
the BB and the Compton cloud would suggest that the system most likely
harbours a NS.
Indeed, the presence of a WD in 4U 2206+54 is basically ruled out because 
of the X--ray spectral shape, which is completely different from that 
observed by e.g. Kubo et al. (1998) and Owens et al. (1999) in the system 
$\gamma$ Cas, which is thought to host a WD.
A better comparison between 4U 2206+54 and systems hosting a WD might
come if we consider magnetic cataclysmic variables, such as intermediate
polars (see e.g. de Martino et al. 2004 and references therein). However,
these objects have spectra with a BB temperature $\approx$30 times lower 
than found in 4U 2206+54, and practically no emission detectable above 30 
keV. Thus, the high BB temperature of 4U 2206+54 is not compatible with 
that of a WD surface, which is expected to mainly emit in the UV rather 
than in the soft X--rays; additionally, the detection of X--ray emission 
up to 60 keV can hardly be explained by assuming a WD as the accretor in 
this system. 

The same line of thought applies to disfavour a BH interpretation: the
Comptonization component has a temperature and an optical depth unusual 
for a BH in its low-hard state and hosted in a MXRB (e.g. Frontera et al. 
2001). So, all the above points to a NS as the accreting object in this 
system, even if no pulsations were ever detected.

Several observed properties of the source are naturally explained by the
accreting NS model. Concerning the X--ray luminosity, accretion onto the
NS from a stellar wind emitted by the O9.5V companion star can easily fit
the observations: following Frank et al. (1992), if we assume that the
companion emits $\approx$10$^{-6}$ $M_\odot$ in the form of a wind, one
needs to hypothesize an accretion efficiency $\eta_{\rm acc} \approx$
10$^{-5}$ to produce a luminosity $L \approx$ 10$^{35}$ erg s$^{-1}$. 
This value of $\eta_{\rm acc}$ may possibly be on the low side
of the allowed values for accretion from stellar wind in close systems;  
however, according to Perna et al. (2003), if corrections to the standard
formulae used to estimate the wind accretion rates are introduced, the
accretion efficiency drops substantially.

Alternatively, as already suggested by Corbet \& Peele (2001), partial
accretion inhibition due to the ``propeller effect" (Illarionov \& Sunyaev
1975; Stella et al. 1986), according to which the magnetosphere of the NS 
acts as a barrier to accretion of matter onto the NS surface, can be at 
work. A fraction of the wind matter can nonetheless flow along the
magnetic field lines and eventually can reach the NS surface.

As regards the secular X--ray variability over a timescale of $\approx$1 
year, we suggest that this might be due to modulations in the wind 
density, such as density waves produced by pulsations of the companion 
star envelope.

In spite of all the above, the NS interpretation rises some problems. In 
particular, as it evidently appears from Table 2, the BB 
radius. Clearly, a size of $\sim$150 m is not acceptable if we assume that 
the whole NS surface is responsible for the BB X--ray emission.
In order to correct for approximations in the BB model application to 
X--ray data, the hardening factor $f$ (Shimura \& Takahara 1995),
defined as the ratio between the color and the effective BB temperatures,
can be introduced. This leads to a corrected BB radius equal to 
$f^{2} \cdot R_{\rm BB}$. However, we should assume that $f \sim$ 8 to 
regain the correct BB size for a NS ($\sim$10 km), while common values for 
$f$ are around 1.7 (Merloni et al. 2000) and extremes do not exceed 
$\sim$3 (Borozdin et al. 1998).
A further possible explanation for the small BB emitting area size,
assuming isotropic emission from the NS surface, is the following:  
because of cooling and back-warming effects the spectrum at the NS
surface, if fitted with a ``classic" BB model, can lead to the net 
result of underestimating the emitting area by as much as 2 orders of
magnitude (London et al. 1986).

The alternative to solve this shortcoming is to assume that the emission 
is not isotropic, i.e. that the accreting matter is either forming a disk 
around the compact object or is funneled onto the NS magnetic polar caps.

The first possibility appears unlikely because accretion
in this system is most probably occurring via stellar wind emitted from
the companion star. This comes from assuming that the 9.6 d periodicity
determined by Corbet \& Peele (2001) is the orbital period of the system
and that the masses of the two components are 1.4 $M_\odot$ for the NS and
$M_{\rm *}$~=~19~$M_\odot$ (Lang 1992) for the O9.5V companion, as
spectroscopically identified by Negueruela \& Reig (2001). With these
values, the Roche lobe radius of the companion is $R_{\rm L} \sim$ 32
$R_\odot$, thus much larger than the radius of a O9.5V star, which is
$R_{\rm *}~\sim$~7.8 $R_\odot$ (Lang 1992). Thus, because the wind has 
very low intrinsic angular momentum, a large accretion disk is unlikely to 
be formed with this accretion mechanism. As remarked in the previous 
Section, the use of a DBB model instead of a BB produces more unstable 
fits to our X--ray dataset.

Alternatively, a magnetically-driven accretion scenario can be considered:  
in this case, one needs the magnetic field of the compact object to be
strong enough to form two accretion columns. Indeed, the tentative
detection of a CRF indicating the presence of a $\approx$10$^{12}$ Gauss
magnetic field (see Orlandini \& Dal Fiume 2001) associated with the NS
would suggest this possibility.

A further indication that the BB emission is indeed anisotropic (i.e., 
confined on a fraction of the NS surface) comes from the estimate of the 
size $r_{\rm seed}$ of the region emitting the Comptonization seed 
photons. Following the prescription by in 't Zand et al. (1999) for the 
computation of $r_{\rm seed}$ and assuming that the Comptonization seed 
photons in 4U 2206+54 are produced by the BB (therefore $kT_{\rm seed}
\sim$ 1.6 keV) we obtain that $r_{\rm seed} \sim$ 0.12 km. This estimate 
is in quite good agreement with our independent determination of the BB 
emission region radius (see Table 2), thus suggesting that indeed the BB 
emitting area covers only a fraction of the NS surface.

However, as no X--ray pulsations are detected from this source, the
magnetically-driven accretion interpretation needs at least one of the
following possibilities to be tenable: an angle between NS spin and
magnetic axes close to zero, or a very low inclination angle for the
system. In this latter case, even if we assume a non-zero but small (e.g.,
few degrees) angle between the NS magnetic field and spin axes, the system
geometry is such that we could continuously see X--ray radiation from a
single polar cap of the NS. Clearly, the emission will not be modulated by
the NS rotation in this case also.

This of course means introducing a fine tuning of the system parameters:  
however, a similar scenario has been proposed to explain the absence of
pulsations from 4U 1700$-$37 (White et al. 1983b) and, more recently, from
4U 1700+24 (Masetti et al. 2002), which are believed to host an accreting
NS. For 4U 1700+24 Galloway et al. (2002) further supported this
description by finding a small amplitude (1 km s$^{-1}$) Doppler
periodicity of $\sim$400 d in their optical spectroscopic data: this
period, earlier suggested by Masetti et al. (2002) from timing analysis of
{\it RXTE}/ASM data, is quite likely produced by the orbital motion of the
system.

A test for the low-inclination hypothesis of 4U 2206+54 can come by
determining, or at least by putting tight constraints on, the orbital
Doppler shift of the companion star:  indeed, assuming the system
parameters discussed above, we find that the orbital velocity of the
companion is v$_{\rm orb} \sim$ 20 km s$^{-1}$; we note that this value is
consistent with the scatter of the system radial velocities measured by
Abt \& Bautz (1963) from optical spectra.

Thus, summarizing, and despite the problems encountered in the analysis of
the observational data on this source, the picture emerging is that 4U
2206+54 is a low-luminosity system composed of a NS and a `normal' blue
main-sequence star; the NS is accreting from the wind coming from the
companion and is orbiting it in a possibly low-inclination orbit.
Tentative evidence of a strong magnetic field from the NS is found, but
deep spectroscopic observations, e.g. with {\it INTEGRAL}, of the hard
X--ray tail of this source are needed to confirm (or disprove) this.
Long-term variations in the X--ray flux from the source can be explained
as due to oscillations in the wind density, possibly induced by slow
pulsations of the companion star envelope. This hypothesis can be tested
with long-term spectrophotometric monitoring of the optical counterpart.

\begin{acknowledgements}

This work has made use of the ASI Science Data Centre Archive at
ESA/ESRIN, Frascati, of the NASA's Astrophysics Data System and of the
SIMBAD database, operated at CDS, Strasbourg, France. {\it BeppoSAX} was a
joint program of Italian (ASI) and Dutch (NIVR) space agencies. This
research has been partially supported by ASI. We are grateful to the 
anonymous referee for his/her comments which helped us to improve this 
paper.

\end{acknowledgements}

\end{document}